\documentclass{webofc}
\usepackage[varg]{txfonts}   % Web of Conferences font
\begin{document}
\title{ Dynamical picture for the exotic XYZ states}
%
% subtitle is optionnal
%
%%%\subtitle{Do you have a subtitle?\\ If so, write it here}

\author{\firstname{Mikhail} \lastname{Ivanov}
\inst{1}\fnsep\thanks{\email{ivanovm@theor.jinr.ru}} 
}

\institute{Bogoliubov Laboratory of Theoretical Physics,
Joint Institute for Nuclear Research,\newline
141980 Dubna, Russia} 

\abstract{
We present a dynamical approach for description of the multi-quark states
that is based on an effective interaction Lagrangian describing the coupling of
hadrons to their constituent quarks. First, we explore the consequences of 
treating the $X(3872)$ meson as a tetraquark bound state. We calculate the decay
widths of the observed channels and conclude that for reasonable values of 
the size parameter of the $X(3872)$ one finds consistency with the available 
experimental data. 
Then we have critically checked the tetraquark picture for 
the $Z_c(3900)$ state by analyzing its strong decays.
We found that  $Z_c(3900)$ has a much more stronger coupling to $DD^\ast$ than
to $J/\psi\pi$ that is in discord with experiment. As an alternative we 
have employed  a molecular-type four-quark current to describe  the decays 
of the  $Z_c(3900)$ state as the charged particle in the isotriplet.
We found that a molecular-type current gives the values of the above
decays  in accordance with the experimental observation.
By using molecular-type four-quark currents for the recently observed 
resonances $Z_b(10610)$ and $Z_b(10650)$, we have calculated their two-body 
decay rates into a bottomonium state plus a 
light meson as well as into B-meson pairs.
}
\maketitle
\section{Introduction}

The most reliable prediction of the quark model is the spectrum of
conventional hadrons composed either from quark-antiquark (mesons)
or from three quarks (baryons). These states have been observed
in experiments and their properties have been carefully studied by
experimentalists as well as theorists for a long time.
However, this situation has been changed since 2003 with the discovery
of many charmonium- and bottomonium-like XYZ states
that do not fit the simple quark-antiquark interpretation.
The specific feature of these states is that their masses are
close to meson-meson thresholds. There are several theoretical interpretations
of them, e.g., as hadronic molecules, tetraquarks, threshold cusps, etc.
(for review, see Refs.~\cite{Lebed:2016hpi,Nielsen:2009uh,Klempt:2007cp}).

The $Y$ states are neutral with quantum numbers $J^{PC}=1^{--}$
of charmonium but not the simple $c\bar c$ charmonium. Among the observed
Y states are $Y(4005)$,  $Y(4260)$,  $Y(4360)$, etc.
The $Z$ states ($Z_c$ and $Z_b$) are basically charged, for example, $Z_c^+$
has  quark content $c\bar c u\bar d$ so it is certainly exotic state.
The $Z_b^+$ has quark content $b\bar b u\bar d$.
The $X$ states are the non-$q\bar q$ mesons but other than the $Y's$ and $Z's$.
The most famous is the $X(3872)$ with quantum numbers $J^{PC}=1^{++}$.

We present a dynamical approach for description of the multi-quark states
that is based on an effective interaction Lagrangian describing the coupling of
hadrons to their constituent quarks. We give a brief sketch of our results
on the decay widths of some exotic four-quark states obtained within this
framework and published in series of our papers of
Refs.~\cite{Dubnicka:2010kz,Dubnicka:2011mm,Goerke:2016hxf,Goerke:2017svb,Gutsche:2017twh,Gutsche:2016cml}. 

First, we explore the consequences of treating the $X(3872)$ meson as a
tetraquark bound state. We calculate the decay widths of the observed channels
and conclude that for reasonable values of the size parameter of the $X(3872)$
one finds consistency with the available experimental data. 
Then we have critically checked the tetraquark picture for 
the $Z_c(3900)$ state by analyzing its strong decays.
We found that  $Z_c(3900)$ has a much more stronger coupling to $DD^\ast$ than
to $J/\psi\pi$ that is in discord with experiment. As an alternative we 
have employed  a molecular-type four-quark current to describe  the decays 
of the  $Z_c(3900)$ state as the charged particle in the isotriplet.
We found that a molecular-type current gives the values of the above
decays  in accordance with the experimental observation.
By using molecular-type four-quark currents for the recently observed 
resonances $Z_b(10610)$ and $Z_b(10650)$, we have calculated their two-body 
decay rates into a bottomonium state plus a 
light meson as well as into B-meson pairs.

\section{Dynamical picture for multiquark states: 
covariant confined quark model}

\begin{figure}[ht]
\centering
\includegraphics[scale=0.4]{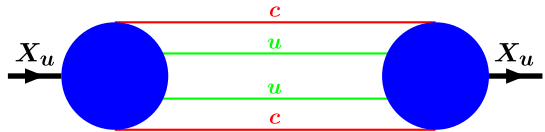} 
\caption{Tetraquark self-energy diagram}
\label{fig:X-mass}       
\end{figure}

The main assumption of approach is that hadrons interact via quark exchange
only. The interaction Lagrangian is written as
\begin{eqnarray*}
{\cal L}_{{\rm int}} &=& g_H \cdot H(x)\cdot J_H(x),
\qquad\qquad\text{where the quark currents are given by}\\[2ex]
 J_M(x) &=& 
\int\!\! dx_1 \!\!\int\!\! dx_2\,
F_M (x;x_1,x_2) \cdot \bar q^a_1(x_1)\, \Gamma_M \,q^a_2(x_2)
 \hspace{2cm} \text{Meson}
\\[2ex]
J_B(x) &=& \int\!\! dx_1 \!\!\int\!\! dx_2 \!\!\int\!\! dx_3\,
F_B (x;x_1,x_2,x_3)  \hspace{3.4cm} \text{Baryon}
\\
&&
\times\, \Gamma_1 \, q^{a_1}_{1}(x_1) \,
\Big( q^{a_2}_{2}(x_2)C \, \Gamma_2 \, q^{a_3}_{3}(x_3)\Big)
\cdot \varepsilon^{a_1a_2a_3}
\\
&&\\
 J^\mu_T(x) &=&
\int\!\! dx_1\ldots\int\!\! dx_4\,
 F_T (x;x_1,\ldots,x_4) \hspace{3.1cm} \text{Tetraquark}
\\
&&
\times\, 
\Big(q_1^{a_1}(x_1)\, C\Gamma_1\, q_2^{a_2}(x_2)
\Big) \cdot
\Big(\bar q_3^{a_3}(x_3)\, \Gamma_2 C\, \bar q_4^{a_4}(x_4)
\Big)\cdot
\varepsilon^{a_1a_2c} \varepsilon^{a_3a_4c}.
\end{eqnarray*}
Here, the $F(x;x_1,\ldots,x_n)$ are the vertex functions characterizing
the distribution of quarks inside the hadron. The matrix elements of
the physical processes are described by the Feynman diagrams which
are convolution of the quark propagators and vertex functions.
The details of all calculations may be found in our published papers of
Refs.~\cite{Dubnicka:2010kz,Dubnicka:2011mm,Goerke:2016hxf,Goerke:2017svb,Gutsche:2017twh,Gutsche:2016cml}. 
For illustration, we show in Fig.~\ref{fig:X-mass} the tetraquark self-energy
diagram which is needed for the determination of coupling constant
$g_T$ in the interaction Lagrangian. 
\section{\boldmath{$X(3872)$}}

A narrow charmonium--like state $X(3872)$ was observed in 2003 in the exclusive
decay process $B^\pm\to K^\pm\pi^+\pi^- J/\psi$ \cite{Choi:2003ue}. 
A mass was found to be very close to the $D^0D^{\ast\,0}$~threshold and 
width less than 2.3~MeV. The state was confirmed in B-decays by the BaBar
experiment~\cite{Aubert:2004fc} 
and in $p\overline{p}$ production by the Tevatron experiments
CDF~\cite{Acosta:2003zx} and D\O{}\cite{Abazov:2004kp}.
From the angular analysis performed by several collaborations
it was shown that the $X(3872)$ state has the quantum number  $J^{PC}=1^{++}$.
Then it was found that the branching ratios of the modes
$X \to \pi^+\pi^- J/\psi$ and $X \to \pi^+\pi^-\pi^0 J/\psi$
are almost the same that imply strong isospin violation.
There are several different interpretations of the $X(3872)$ in the literature:
a molecule bound state ($D^0\overline{D}^{\ast\,0}$),  threshold cusps,
hybrids and glueballs.

We employed the  diquark-antidiquark interpretation suggested in
Ref.~\cite{Maiani:2004vq} and performed  an independent analysis of 
the properties of the $X(3872)$ within our covariant confined quark model
in the papers Refs.~\cite{Dubnicka:2010kz,Dubnicka:2011mm}. 
The quark content was chosen as  
$[cq]_{S=0}\,[\bar c \bar q]_{S=1} + [cq]_{S=1}\,[\bar c \bar q]_{S=0}$,
$(q=u,d)$~\cite{Maiani:2004vq} .
The physical states are suppossed to be a linear
superposition of the $X_u$ and $X_d$ states according to
\[
X_l\equiv X_{\rm low} =  X_u\, \cos\theta +  X_d\, \sin\theta,
\qquad
X_h\equiv X_{\rm high} = - X_u\, \sin\theta +  X_d\, \cos\theta.
\]
The mixing angle $\theta$ is determined from fitting the ratio
of branching ratios of the $X-$decays into $J/\psi+2\pi$ and  $J/\psi+3\pi$.

\begin{figure}[ht]
\centering
\begin{tabular}{lr}
\includegraphics[scale=0.3]{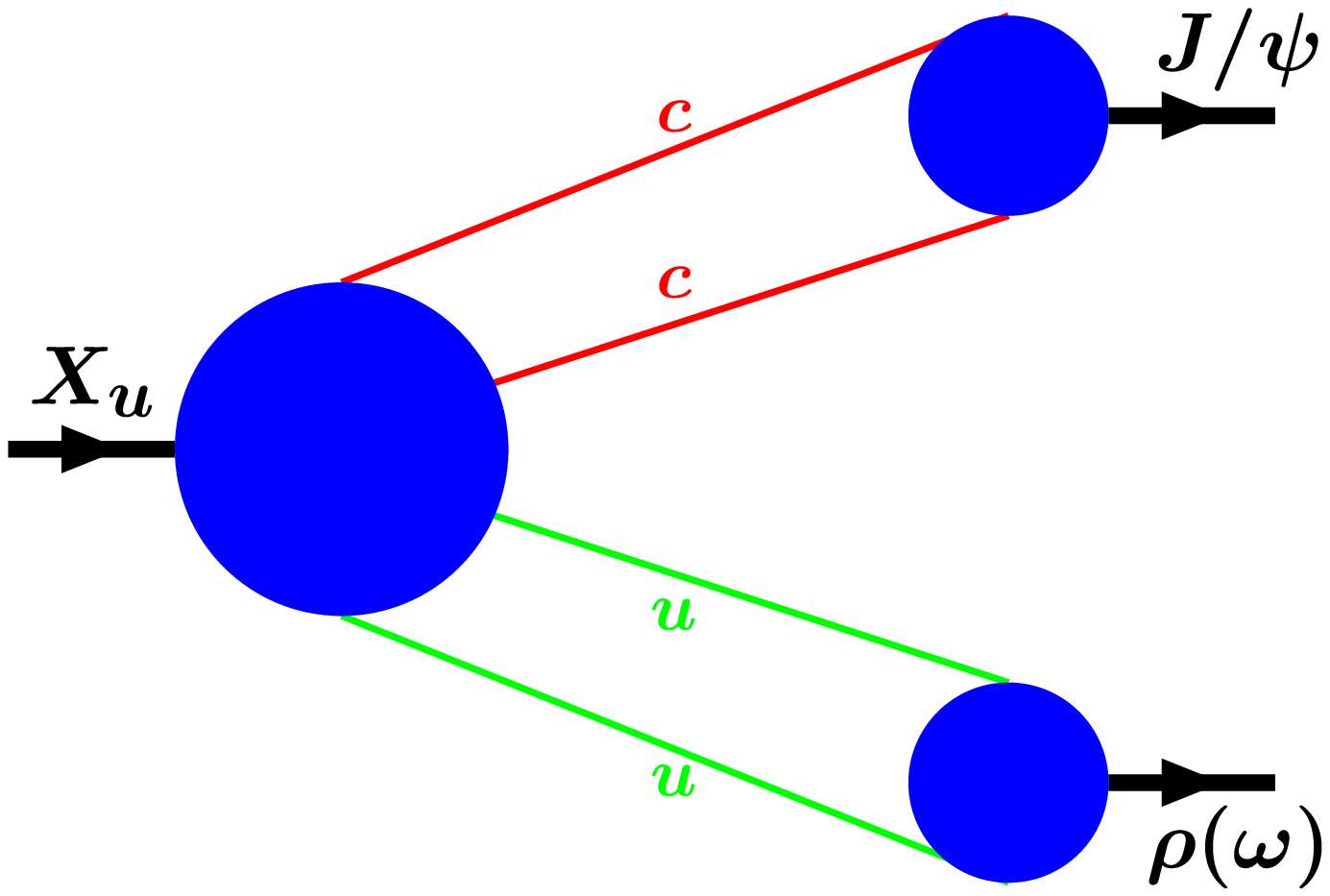} &
\includegraphics[scale=0.3]{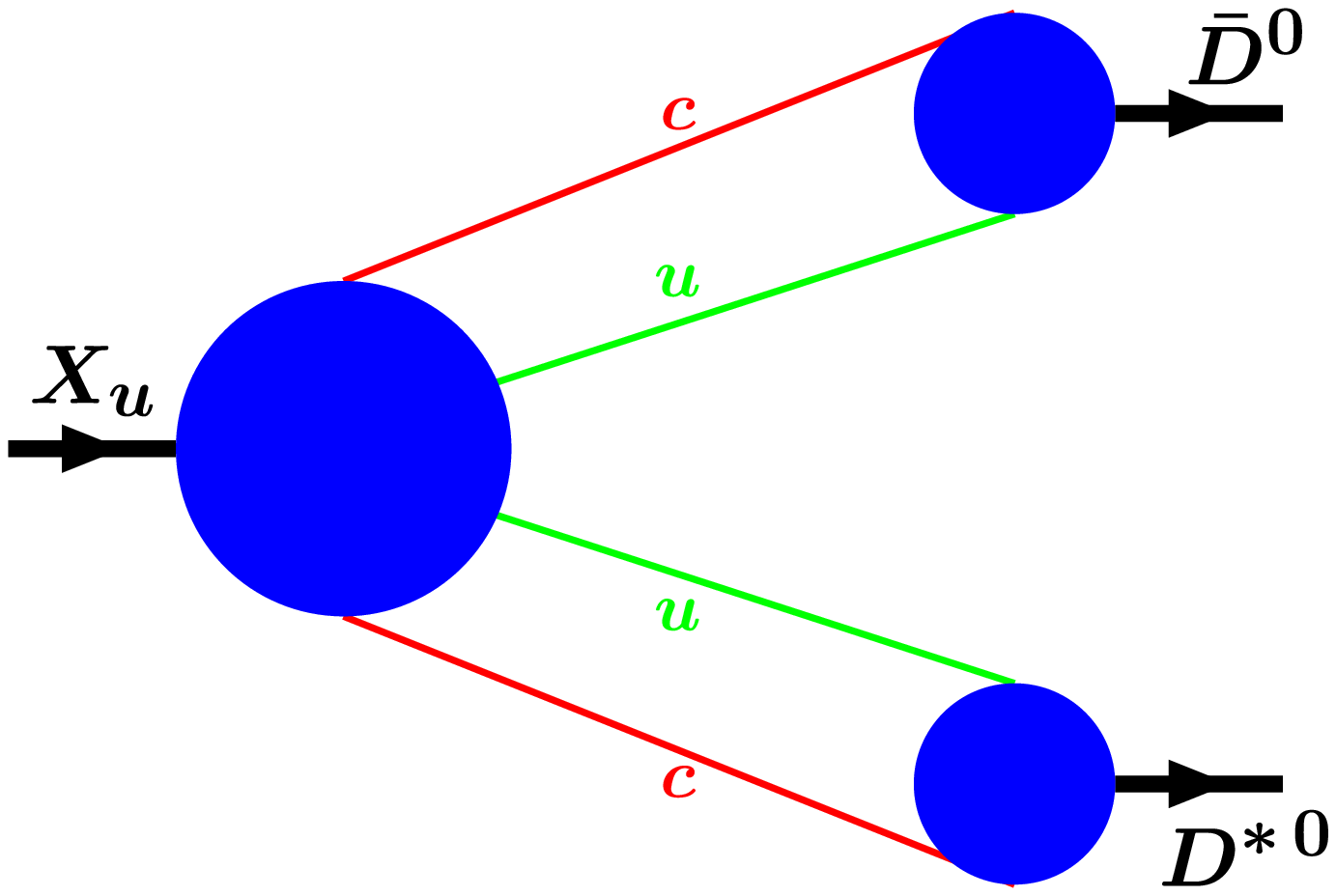}
\end{tabular}
\caption{Diagrams describing the strong X-decays}
\label{fig:X-strong}     
\end{figure}

\begin{figure}[ht]
\centering
\begin{tabular}{lr}
\includegraphics[scale=0.32]{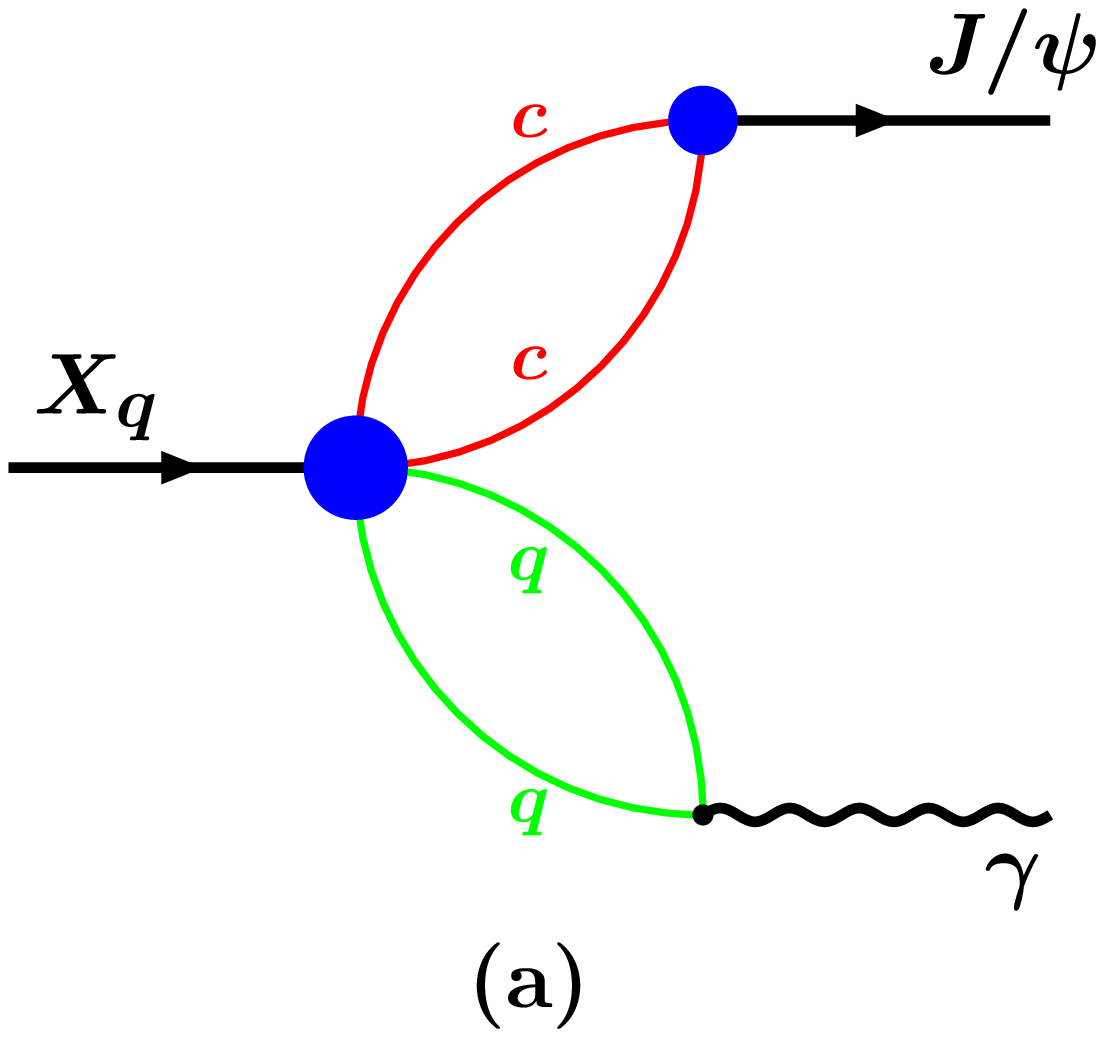} & 
\includegraphics[scale=0.32]{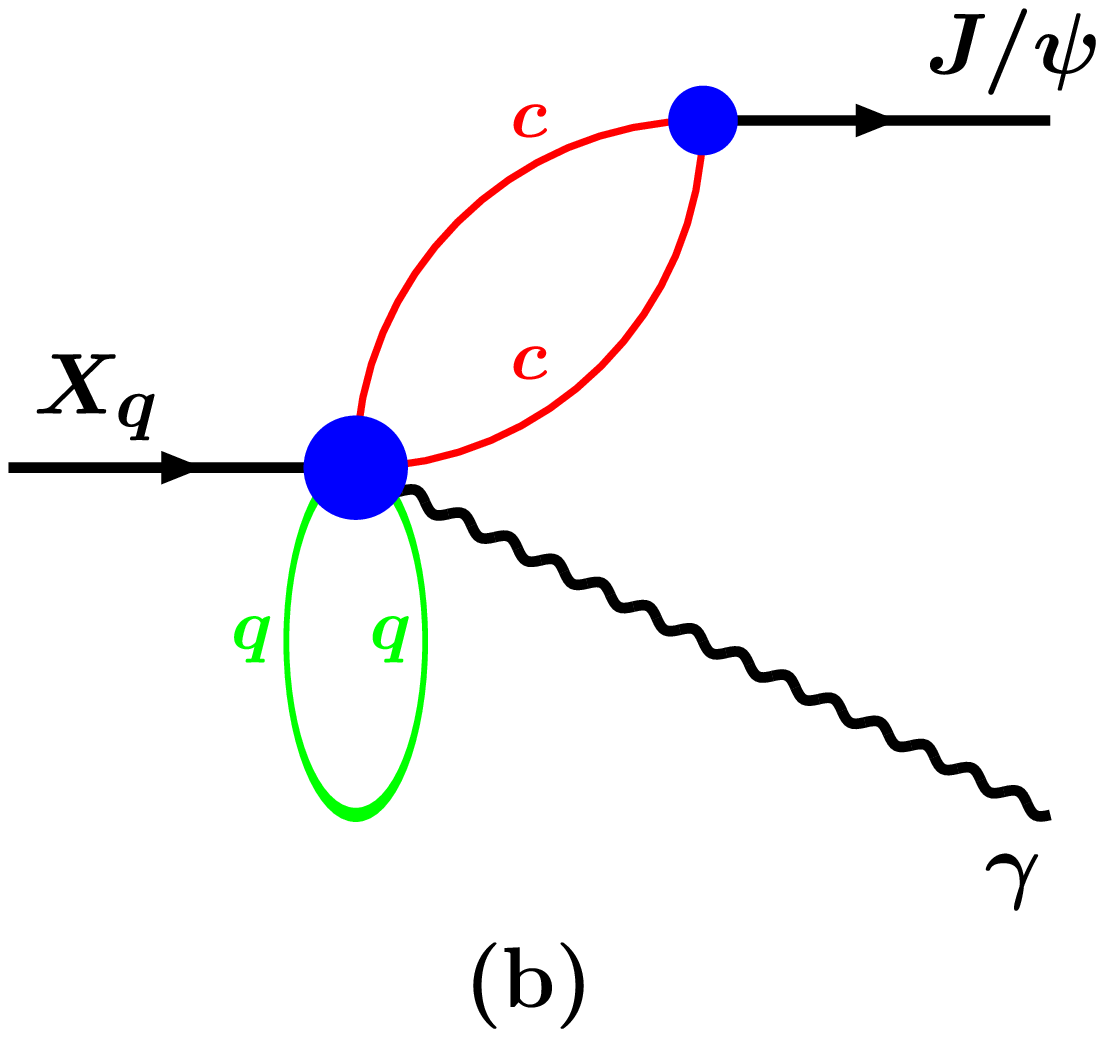}\\[2ex]
\includegraphics[scale=0.32]{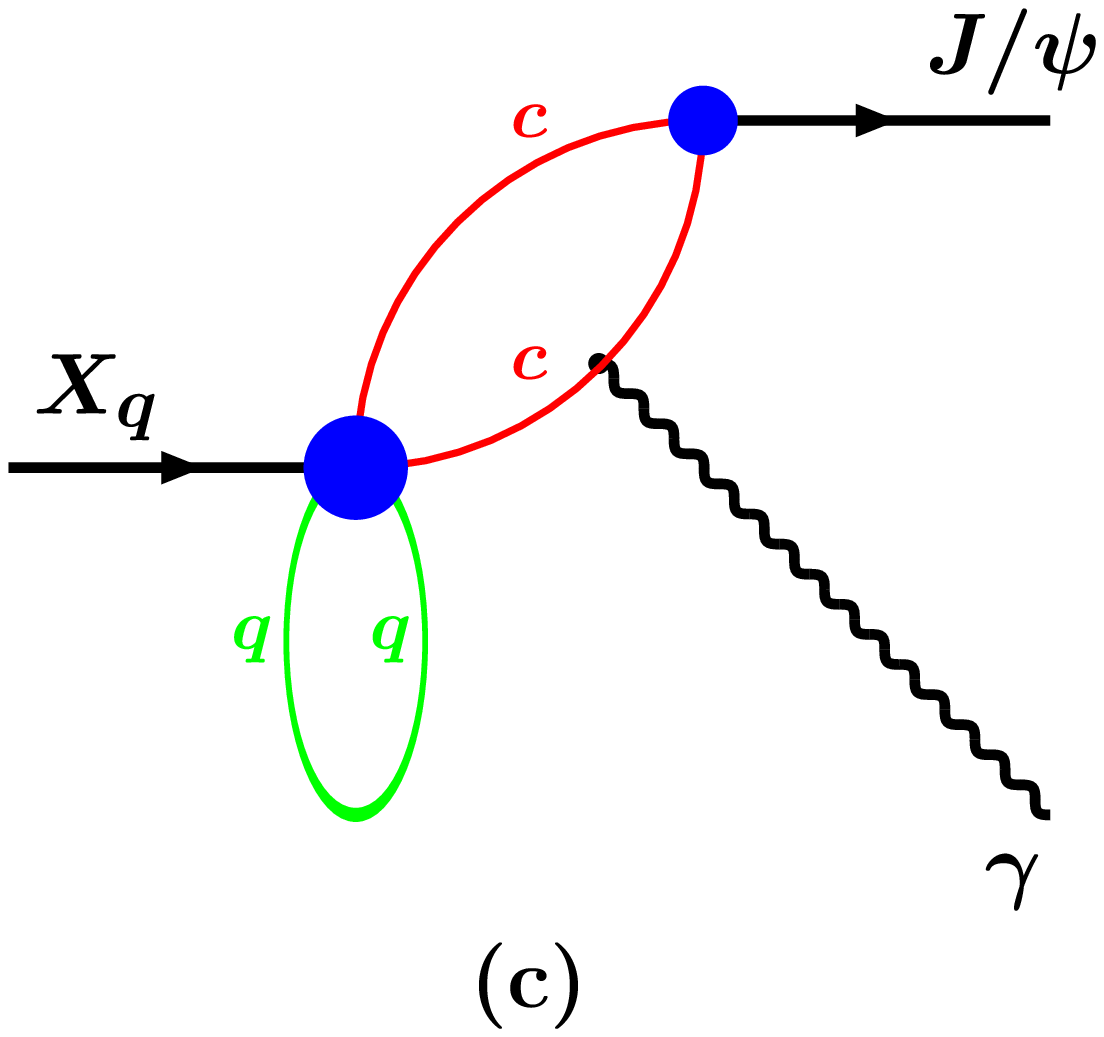} & 
\includegraphics[scale=0.32]{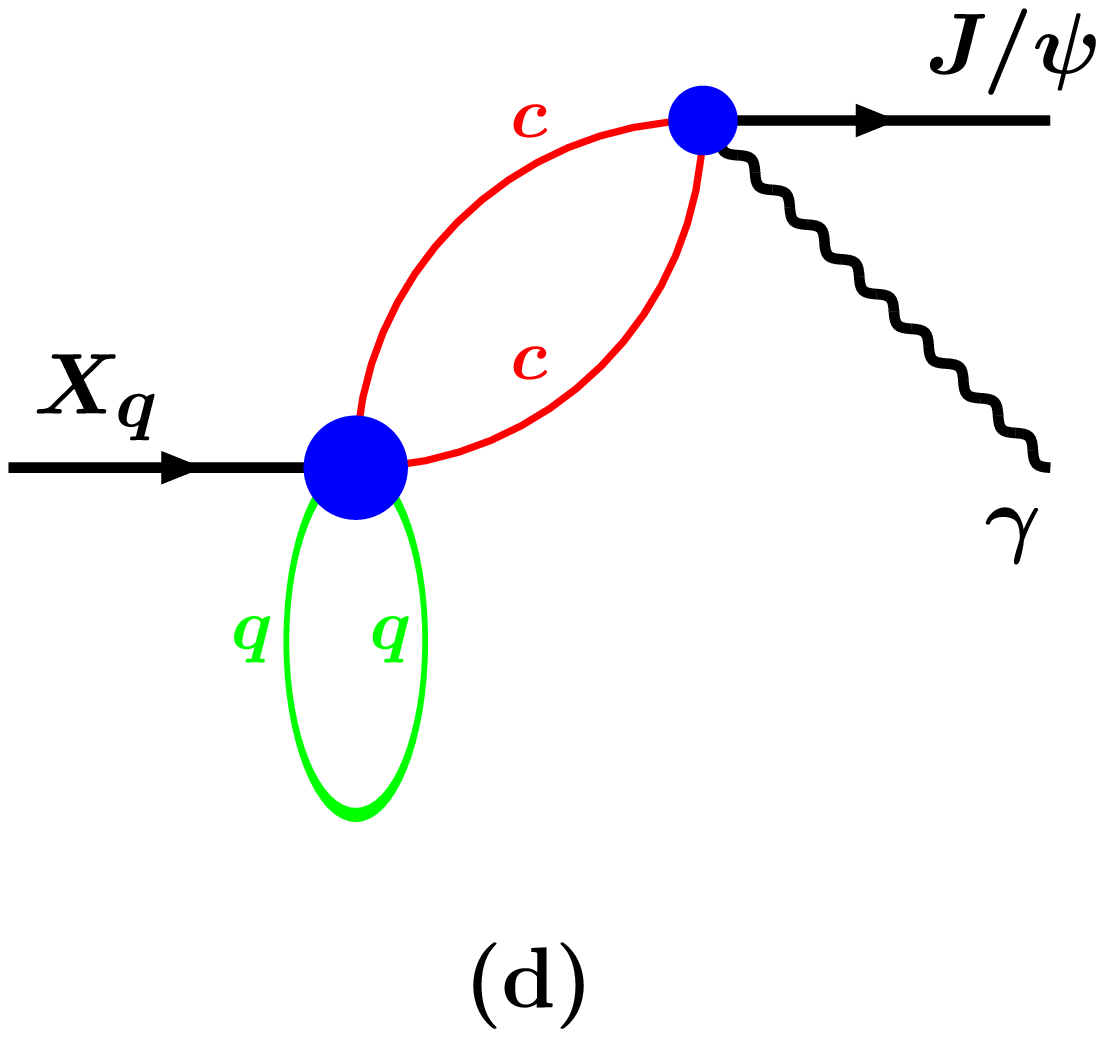}
\end{tabular}
\caption{Diagrams describing the radiative X-decays}
\label{fig:X-rad}
\end{figure}

The diagrams describing the strong decays $\Gamma(X\to J/\psi+n\,\pi)$  (n=2,3)
and $X\to D^0\bar D^0\pi^0 $ are shown in Fig.~\ref{fig:X-strong}.
The diagrams describing the radiative decays 
$\Gamma(X\to J/\psi+\gamma)$  are shown in Fig.~\ref{fig:X-rad}.

\begin{figure}[ht]
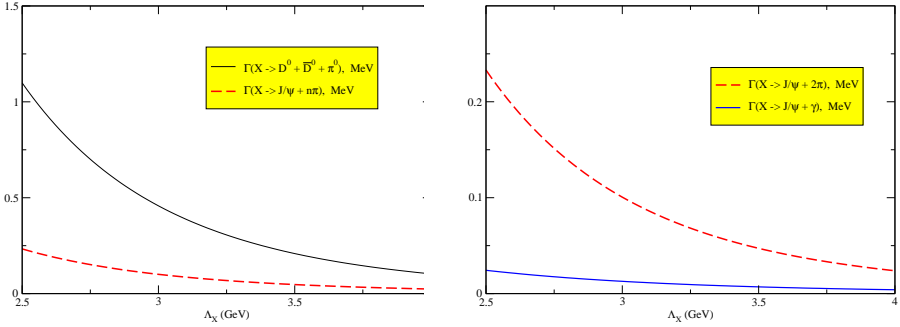

\centering
\begin{tabular}{lr}
\includegraphics[scale=0.25]{fig_width-X} &
\includegraphics[scale=0.25]{fig_X_elem}
\end{tabular}
\caption{Dependence of the strong (left panel) and  radiative (right panel)
decay widths onthe size parameter}
\label{fig:X-width}       
\end{figure}

The only free parameter is the so-called  size parameter $\Lambda_X$
which appears in the vertex function $F_T(x;x_1,\ldots,x_4)$.
We plot  the dependence of the calculated decay widths on this parameter
in Fig.~\ref{fig:X-width} in quite large range of $\Lambda_X\in [2.5-4]$~GeV.

Finally, we compare the obtained results for the ratios of decay widths
with the available experimental data. One can see that there is reasonable
agreement in the case of the strong decays and very good agreement
in the case of the radiative decays.
\[
\frac{\Gamma(X\to D^0\bar D^0 \pi^0)}
     {\Gamma(X\to J/\psi\pi^+\pi^-)}  = 
\left\{\begin{array}{lr} 4.5 \mbox{ $\pm$} 0.2 & \mbox{theor} \\[1.5ex]
                         10.5 \mbox{$\pm$} 4.7 & \mbox{expt}
        \end{array} \right|,
\qquad
\frac{\Gamma(X\to J/\psi + \gamma)}
     {\Gamma(X\to J/\psi+2\pi)} =
\left\{\begin{array}{lr} 0.15 \mbox{ $\pm$} 0.03 & \mbox{theor} \\[1.5ex]
                         0.14 \mbox{$\pm$}  0.05 & \mbox{expt}
        \end{array} \right|.
\]
%
%----------------------------------------------------------
%
\section{\boldmath{$Z_c(3900)$} }

The BESIII Collaboration \cite{Ablikim:2013mio} observed a structure
in the  $\pi^\pm J/\psi$ mass spectrum of the process
$ e^+e^-\to \pi^+ \underbrace{\pi^-J/\psi}_{Z_c^-} $.
It was named as the  $Z_c(3900)$ state with a mass near 3.9~GeV
and a width around of 50~MeV. This state was confirmed by
the Belle Collaboration~\cite{Liu:2013dau}.
Recently, the D0 Collaboration presented evidence for
the exotic charged $Z_c(3900)$ in semi-inclusive weak decays of $b$-flavored
hadrons \cite{Abazov:2018cyu}.
This structure can be interpreted as a new charged charmonium-like state. 
Mode with $D\bar D^\ast$ in the final state  was studied by
the BESIII Collaboration  \cite{Ablikim:2013xfr}.  A distinct charged structure 
was observed in the $ (D\bar D^\ast)^\mp $ invariant mass distribution
of the process $ e^+e^-\to \pi^\pm \underbrace{(D\bar D^\ast)^\mp}_{Z_c^\mp}  $.
The angular distribution of the $\pi Z_c(3885)$  system favors 
a $J^P = 1^+$ quantum number assignment for the new structure.
The ratio of partial widths was determined as 
\begin{equation}
\frac{\Gamma(Z_c(3885)\to D\bar D^\ast)}{\Gamma(Z_c(3900)\to\pi J/\psi)}
=6.2\pm 1.1 \pm 2.7
\end{equation}

Assuming that the $Z_c$ the charged partner of the $X(3872)$ state
one can write down a tetraquark-type current: 
\begin{equation}
J^\mu = \frac{i}{\sqrt{2}}\varepsilon_{abc}\varepsilon_{dec}
\left[  (u^T_a C\gamma_5 c_b)(\bar d_d\gamma^\mu C \bar c^T_e)
      - (u^T_a C\gamma^\mu c_b)(\bar d_d\gamma_5 C \bar c^T_e)\right]
\end{equation}
In Ref.~\cite{Goerke:2016hxf} we  have used the nonlocal generalization
of this tetraquark current to calculate a number of the $Z_c(399)$
two-body decay widths of the process $1^+(p,\mu) \to 1^-(q_1,\nu)+ 0^-(q_2)$.
It was found  that, in our model, the leading Lorentz metric structure in the
matrix elements describing the decays $Z_c^(3900)\to\bar DD^{\ast}$ vanishes
analytically. This results in a significant suppression of these decay widths by
the smallness of the relevant phase space factor ${\bf|q|}^5$. 
Since the experimental data \cite{Ablikim:2013xfr} show
that the $Z_c(3900)$ has a much more stronger coupling to $DD^\ast$ than
to $J/\psi\pi$, one has to conclude that the tetraquark-type current  
for the $Z_c(3900)$ is in discord with experiment.

As an alternative we have employed  a molecular-type four-quark current
to describe the decays of the $Z_c(3900)$ state:
\begin{equation} 
J^\mu = \frac{1}{\sqrt{2}} 
\left[ (\bar d \gamma_5 c) (\bar c\gamma^\mu u)
      +(\bar d \gamma^\mu c)(\bar c\gamma_5  u) \right].
\end{equation}
In this case we found that for a relatively large size parameter
$\Lambda_{Z_c} \sim 3.3$~GeV 
one can obtain the partial widths of the decays $Z_c^+(3900)\to\bar DD^{\ast}$ 
at the order  $\sim 15$~MeV for each mode. At the same time
the partial widths for decays $Z_c^+(3900)\to J/\psi\pi^+\,, \eta_c\rho^+$
are suppressed by a factor of $6-7$ in accordance with experimental
data~\cite{Ablikim:2013xfr}.
If the $\Lambda_{Z_c}$ is varied in the limits  $\Lambda_{Z_c}=3.3 \pm 1.1$~GeV
then one has 
\begin{eqnarray*}
  \Gamma(Z^+_c\to J/\psi+\pi^+) &=& (1.8 \pm 0.3)\,\text{MeV}\,,
  \hspace*{0.7cm}
  \Gamma(Z^+_c\to \bar D^0 + D^{\ast\, +}) = (10.0^{+1.7}_{-1.4})\,\text{MeV}\,,
\nonumber\\[1.2ex]
\Gamma(Z^+_c\to\eta_c+\rho^+) &=& (3.2^{+0.5}_{-0.4})\,\text{MeV}\,,
 \hspace*{1.05cm}
\Gamma(Z^+_c\to \bar D^{\ast\, 0} + D^+) = (9.0^{+1.6}_{-1.3})\,\text{MeV}\,.
\end{eqnarray*}
Preliminary data from BESIII cited in \cite{Yuan:2018inv} 
were reported for the ratio
\begin{equation}
R(Z) = \frac{\mathcal{B}(Z_c(3900)\to\rho\eta_c)}
            {\mathcal{B}(Z_c(3900)\to\pi J/\psi)} = 2.1 \pm 0.8.
\end{equation}
They agree very well with our finding $R(Z) = 1.8 \pm 0.4$. 

%---------------------------------------------------------------------

\section{\boldmath{$Z_b(10610)$} and \boldmath{$Z'_b(10610)$} }

A few years ago the Belle Collaboration~\cite{Belle:2011aa}
reported on the observation of two charged bottomoniumlike resonances
in the mass spectra of $\pi^{\pm}\Upsilon(nS)$ ($n=1,2,3$) and
$\pi^{\pm}h_b(mP)$ ($m=1,2$) in the decays 
$\Upsilon(5S)\to\Upsilon(nS)\pi^+\pi^-, h_b(mP)\pi^+\pi^-$. 
The measured masses and widths were given by
\begin{eqnarray*}
      M_{Z_b} &=& (10607.2 \pm 2.0 ) \ \text{MeV}\,, \qquad  
  \Gamma_{Z_b} = ( 18.4 \pm 2.4 ) \ \mathrm{MeV}\,,
\nonumber\\
M_{Z'_b} &=& ( 10652.2 \pm 1.5) \ \mathrm{MeV}\,, \qquad 
\Gamma_{Z'_b} = ( 11.5 \pm 2.2 ) \ \mathrm{MeV}\,.
\label{eq:Belle-1} 
\end{eqnarray*}
and the quantum numbers are $I^G(J^P)=1^+(1^+)$.
The existence of these two states was later confirmed by the same
collaboration~\cite{Garmash:2014dhx,Garmash:2015rfd} in differing
decay channels.
In the paper~\cite{Garmash:2015rfd} the Belle Collaboration reported on
the results of an analysis of the three-body processes
$e^+e^-\to B\bar B\pi^\pm, B\bar B^\ast\pi^\pm$, and 
$ B^\ast\bar B^\ast\pi^\pm$.
It was found that the transitions $Z^\pm_b(10610)\to [B\bar B^\ast + c.c.]^\pm$
and  
$Z^\pm_b(10650)\to [B^\ast\bar B^\ast]^\pm$ dominate among
the corresponding final states.

Since the masses of the  $Z^+_b(10610)$ and 
$Z_b^\prime(10650)$ 
are very close to the respective $B^\ast\bar B$ (10604 MeV)
and  $B^\ast\bar B^\ast$ (10649 MeV) thresholds, 
it was suggested in  Ref.~\cite{Bondar:2011ev} that they have
molecular-type binding structures. 
\begin{equation}
J^\mu_{Z_b^+} = \frac{1}{\sqrt{2}} 
\left[ (\bar d \gamma_5 b) (\bar b  \gamma^\mu u)
      +(\bar d \gamma^\mu b)(\bar b \gamma_5  u) \right]\,, 
\qquad
J^{\mu\nu}_{Z_b^{\prime +}} = \varepsilon^{\mu\nu\alpha\beta} 
(\bar d \gamma_\alpha b) (\bar b \gamma_\beta u).
\end{equation}
Such a choice guarantees that the  $Z_b$-state can only decay 
to the $[\bar B^\ast B+c.c.]$ pair whereas  
the $Z'_b$-state can decay only to 
a $\bar B^\ast B^\ast$ pair. 
Decays into the $BB$-channels are forbidden.  
Due to $G$-parity  conservation
$Z_b\to \Upsilon+\rho$, $ Z_b\to      \eta_b + \pi$, 
$  Z_b\to      \chi_{b1} + \pi$,
$  Z_b\to    h_b + \rho$.
The decay  $Z_b\to \chi_{b1} + \rho$ is not allowed 
kinematically.

There are therefore only the three allowed decays: 
$Z^+_b \to \Upsilon+\pi^+$, 
$Z^+_b  \to h_b+\pi^+$ and 
$Z^+_b  \to \eta_b+\rho^+$.
The only two new parameters are the size parameters of the two exotic
$Z_b(Z'_b)$ states. As a guide to adjust them  
we take the experimental values of
the largest branching fractions presented by Belle:
\begin{equation}
{\cal B}(Z_b^+\to [B^+ \bar B^{\ast\,0} + \bar B^0 B^{\ast\,+}]) 
= 85.6^{+1.5+1.5}_{-2.0-2.1}\,\%\,,
\qquad    
    {\cal B}(Z_b^{\prime +} \to \bar B^{\ast\,+} B^{\ast\,0})
    = 73.7^{+3.4+2.7}_{-4.4-3.5}\,\%\,.
\end{equation}

By using the central values of these branching rates and 
total decay widths we find the central values of our
size parameters $\Lambda_{Z_b} =3.45$~GeV and  
                $\Lambda_{Z'_b}=3.00$~GeV.
Allowing them to vary in the interval
$ \Lambda_{Z_b} =3.45\pm 0.05 $~GeV and
$ \Lambda_{Z'_b}=3.00\pm 0.05 $~GeV
we obtain the values of various decay widths.
%
%----------------- Tables -----------------------------
%
\begin{table}
\centering
\caption{$Z_b(10610)$ and $Z'_b(10610)$: numerical results}
\label{tab-1}       
\def\arraystretch{1.5}
\begin{tabular}{lll}
\hline
  ~Channel~\hspace*{5mm}  & \multicolumn{2}{c}{Widths, MeV}   \\
              & \qquad $Z_b(10610)$ \qquad  & \qquad  $Z'_b(10650)$ \qquad      
\\
\hline 
 $\Upsilon(1S)\pi^+$ & \qquad  $5.9\pm 0.4$   
                     & \qquad  $9.5^{+0.7}_{-0.6}$ 
\\
 $h_b(1P)\pi^+$ & \qquad  $(0.14\pm 0.01)\cdot 10^{-1}$
               & \qquad  $0.74^{+0.05}_{-0.04}\cdot 10^{-3}$
\\
 $\eta_b \rho^+$  & \qquad   $4.4\pm 0.3$ 
                  & \qquad  $7.5^{+0.6}_{-0.5}$
\\
$B^+\bar{B}^{*0}+\bar{B}^0B^{*+}$ &  \qquad  $20.7^{+1.6}_{-1.5}$   
                                & \qquad   $-$       
\\
 $B^{*+}\bar{B}^{*0}$              &  \qquad  $-$ 
                                 & \qquad  $17.1^{+1.5}_{-1.4}$ 
\\
\hline
  \end{tabular}
%\vspace*{5cm}  
\end{table}
\begin{table}
\centering
\caption{$Z_b(10610)$ and $Z'_b(10610)$: Total widths, MeV  }
\label{tab-2}
\def\arraystretch{1.5}
  \begin{tabular}{lll}  
\hline
 &  \qquad Theory  &  \qquad Belle Expt.  \\      
\hline
$Z_b(10610)$  &  \qquad $30.9^{+2.3}_{-2.1}$
              &  \qquad $25 \pm 7$\\
$Z'_b(10650)$ &  \qquad $34.1^{+2.8}_{-2.5}$ 
              &  \qquad $ 23 \pm 8$    
\\
\hline
  \end{tabular}
%\vspace*{5cm}  
\end{table}

\section{Acknowledgments}

I am grateful to the organizers of International Workshop on QCD,
theory and experiment (QCD$@$Work) for invitation and
hospitality in Matera.
This work was supported in part by the Joint Research Project of Institute
of Physics, SAS and Bogoliubov Laboratory of Theoretical Physics,
JINR, No. 01-3-1114.

\end{document}